\documentclass[sigconf]{acmart} %

\usepackage{booktabs}
\usepackage{graphicx, epstopdf}
\usepackage{subfigure}
\usepackage{url}
\usepackage{multirow}
\usepackage{algorithm}
\usepackage{algorithmic}
\usepackage{balance}
\usepackage{arydshln}

\renewcommand*{\textrightarrow}{$\rightarrow$}

\begin{document}

\title{Masked Multi-Domain Network: Multi-Type and Multi-Scenario Conversion Rate Prediction with a Single Model}
\author{Wentao Ouyang}
\author{Xiuwu Zhang}
\affiliation{%
  \institution{Alibaba Group}
}
\email{maiwei.oywt@alibaba-inc.com}
\email{xiuwu.zxw@alibaba-inc.com}

\author{Chaofeng Guo}
\author{Shukui Ren}
\affiliation{%
  \institution{Alibaba Group}
}
\email{chaofeng.gcf@alibaba-inc.com}
\email{shukui.rsk@alibaba-inc.com}

\author{Yupei Sui}
\author{Kun Zhang}
\affiliation{%
  \institution{Alibaba Group}
}
\email{yupei.syp@alibaba-inc.com}
\email{jerry.zk@alibaba-inc.com}

\author{Jinmei Luo}
\author{Yunfeng Chen}
\affiliation{%
  \institution{Alibaba Group}
}
\email{cathy.jm@alibaba-inc.com}
\email{chenyunfeng@alibaba-inc.com}

\author{Dongbo Xu}
\author{Xiangzheng Liu}
\affiliation{%
  \institution{Alibaba Group}
}
\email{dongbo.xdb@alibaba-inc.com}
\email{xiangzheng.lxz@alibaba-inc.com}

\author{Yanlong Du}
\affiliation{%
  \institution{Alibaba Group}
}
\email{yanlong.dyl@alibaba-inc.com}

\renewcommand{\shortauthors}{Wentao Ouyang et al.}
\renewcommand{\shorttitle}{Masked Multi-Domain Network}

\begin{abstract}
In real-world advertising systems, conversions have different types in nature and ads can be shown in different display scenarios, both of which highly impact the actual conversion rate (CVR). This results in the multi-type and multi-scenario CVR prediction problem. A desired model for this problem should satisfy the following requirements: 1) Accuracy: the model should achieve fine-grained accuracy with respect to any conversion type in any display scenario. 2) Scalability: the model parameter size should be affordable. 3) Convenience: the model should not require a large amount of effort in data partitioning, subset processing and separate storage.

Existing approaches cannot simultaneously satisfy these requirements. For example, building a separate model for each (conversion type, display scenario) pair is neither scalable nor convenient. Building a unified model trained on all the data with conversion type and display scenario included as two features is not accurate enough.

In this paper, we propose the Masked Multi-domain Network (MMN) to solve this problem. To achieve the accuracy requirement, we model domain-specific parameters and propose a dynamically weighted loss to account for the loss scale imbalance issue within each mini-batch. To achieve the scalability requirement, we propose a parameter sharing and composition strategy to reduce model parameters from a product space to a sum space. To achieve the convenience requirement, we propose an auto-masking strategy which can take mixed data from all the domains as input. It avoids the overhead caused by data partitioning, individual processing and separate storage. Both offline and online experimental results validate the superiority of MMN for multi-type and multi-scenario CVR prediction. MMN is now the serving model for real-time CVR prediction in UC Toutiao.
\end{abstract}

%\ccsdesc[500]{Information systems~Computational advertising}


\copyrightyear{2023}
\acmYear{2023}
\setcopyright{acmlicensed}\acmConference[CIKM '23]{Proceedings of the 32nd ACM International Conference on Information and Knowledge Management}{October 21--25, 2023}{Birmingham, United Kingdom}
\acmBooktitle{Proceedings of the 32nd ACM International Conference on Information and Knowledge Management (CIKM '23), October 21--25, 2023, Birmingham, United Kingdom}
\acmPrice{15.00}
\acmDOI{10.1145/3583780.3614697}
\acmISBN{979-8-4007-0124-5/23/10}

\settopmatter{printacmref=true}

\maketitle

\section{Introduction}
Conversion rate (CVR) prediction \cite{lee2012estimating,chapelle2014modeling,lu2017practical} is an essential task in online advertising systems. In optimized Cost-Per-Click (oCPC)
and Cost-Per-Action (CPA) advertising, advertisers seek to maximize conversions for a given budget \cite{zhu2017optimized,ma2018entire,pan2019predicting}.
The predicted CVR impacts both the ad ranking strategy and the ad charging model.

Conversions have different types in nature (e.g., \emph{purchase} a product vs. \emph{sign up} an account) and ads can be shown in different display scenarios (e.g, shown as a \emph{banner} on a page vs. interleaved with news in the \emph{news channel}), both of which highly impact the actual CVR.
As an evidence, the ground-truth CVRs vary significantly over different conversion types (e.g., from 6.3\% to 18.9\%) and over different display scenarios (e.g., from 0.9\% to 14.9\%) in the Criteo dataset (Table \ref{tab_stat}). This results in the multi-type and multi-scenario CVR prediction problem.

A desired model for this problem should satisfy the following requirements: 1) Accuracy: the model should achieve fine-grained accuracy with respect to any conversion type in any display scenario. 2) Scalability: the model parameter size should be affordable. 3) Convenience: the model should not require a large amount of effort in data partitioning, subset processing and separate storage.

One approach to solve this problem is to build a separate model for each (conversion type, display scenario) pair. However, it is neither scalable nor convenient. Another approach is to build a unified model trained on all the data, with conversion type and display scenario included as two feature fields. However, it fails to capture different natures of various conversion types and display scenarios, and is thus not accurate enough.

\begin{figure}[!t]
\centering
\includegraphics[width=0.42\textwidth, trim = 0 0 0 0, clip]{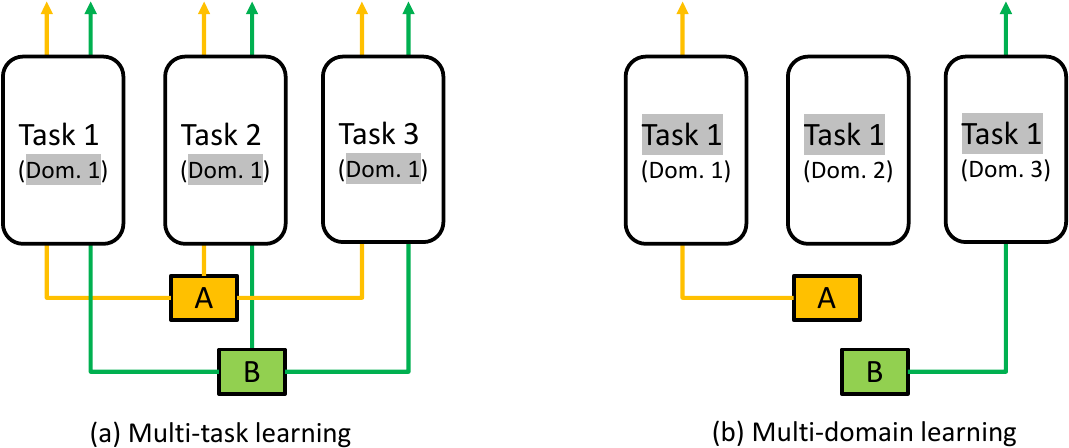}
\vskip -10pt
\caption{$A$ and $B$ are two input instances. (a) In multi-task learning, domains are the same and tasks are different. One input impacts the model parameters of all tasks. (b) In multi-domain learning, domains are different and tasks are the same. One input impacts the parameters for that domain.}
\vskip -10pt
\label{fig_diff}
\end{figure}

In this paper, we formulate the multi-type and multi-scenario CVR prediction problem as a multi-domain learning problem. We would like to illustrate the differences between multi-task learning \cite{caruana1997multitask} and multi-domain learning \cite{dredze2010multi} in Figure \ref{fig_diff}. In multi-task learning, domains are the same and tasks are different. For example, given a video, simultaneously estimating what rating a user will give and how long the user will spend on the video is a multi-task learning problem \cite{zhao2019recommending}. In multi-domain learning, domains are different and tasks are the same. For example, estimating the CVRs of two instances of conversion type ``purchase'' and ``sign up'' by the same model is a multi-domain learning problem.
Moreover, in multi-task learning, one input impacts the model parameters of all tasks; while in multi-domain learning, one input only impacts the model parameters corresponding to that domain.
Because of such differences, existing multi-task learning models such as Shared-Bottom, OMoE, MMoE, CGC and PLE \cite{ma2018modeling,zhao2019recommending,ma2019snr,tang2020progressive} cannot well address our problem.
MT-FwFM \cite{pan2019predicting}, STAR \cite{sheng2021one} and ADIN \cite{jiang2022adaptive} are more relevant state-of-the-art multi-domain models. However, as they are not designed for our problem, they also cannot satisfy the aforementioned three requirements.

Given these limitations, we propose the Masked Multi-domain Network (MMN) in this paper. We will show how we design various strategies to make MMN simultaneously satisfy the accuracy, scalability and convenience requirements.

The main contributions of this work are summarized as follows:
\begin{enumerate}
\item We propose the MMN model to address the multi-type and multi-scenario CVR prediction problem in a real-world advertising system.
\item In order to achieve the accuracy requirement, we model domain-specific parameters and we propose a dynamically weighted loss to account for the loss scale imbalance issue within each mini-batch.
\item In order to achieve the scalability requirement, we propose a parameter sharing and composition strategy to reduce model parameters from a product space to a sum space.
\item In order to achieve the convenience requirement, we propose an auto-masking strategy which can take mixed data from all the domains as input. It avoids the overhead caused by data partitioning, individual processing and separate storage.
\item We conduct both offline and online experiments to test the performance of MMN and several state-of-the-art methods for multi-type and multi-scenario CVR prediction.
\end{enumerate}

\section{Masked Multi-domain Network}
\subsection{Multi-Type and Multi-Scenario CVR Prediction}
In omnimedia advertising platforms, different conversion types (e.g., purchase a product, sign up an account or fill out an online form) and different display scenarios (e.g., shown as a banner, interleaved with news or interleaved with micro-videos) highly impact the actual CVR.
In the Criteo dataset (Table \ref{tab_stat}), the ground-truth CVRs vary significantly over different conversion types (e.g., from 6.3\% to 18.9\%) and over different display scenarios (e.g., from 0.9\% to 14.9\%). In our news feed advertising dataset (Table \ref{tab_stat}), the corresponding values vary from 0.05\% to 27.6\% and from 1.6\% to 4.1\% respectively.

Denote the input instance as $x$, which contains multiple features, such as user features, ad features, context features and cross features.
Denote the conversion type of the ad as $f_t(x)$ and the display scenario of the ad as $f_s(x)$.
Further denote the click label as $y \in \{0,1\}$ and the conversion label as $z \in \{0, 1\}$.
Mathematically, the multi-type and multi-scenario CVR prediction problem is to estimate $p(z = 1 | y = 1, x, f_t(x), f_s(x))$. It is a feature-rich problem.

\begin{figure*}[!t]
\centering
\includegraphics[width=0.7\textwidth, trim = 0 0 0 0, clip]{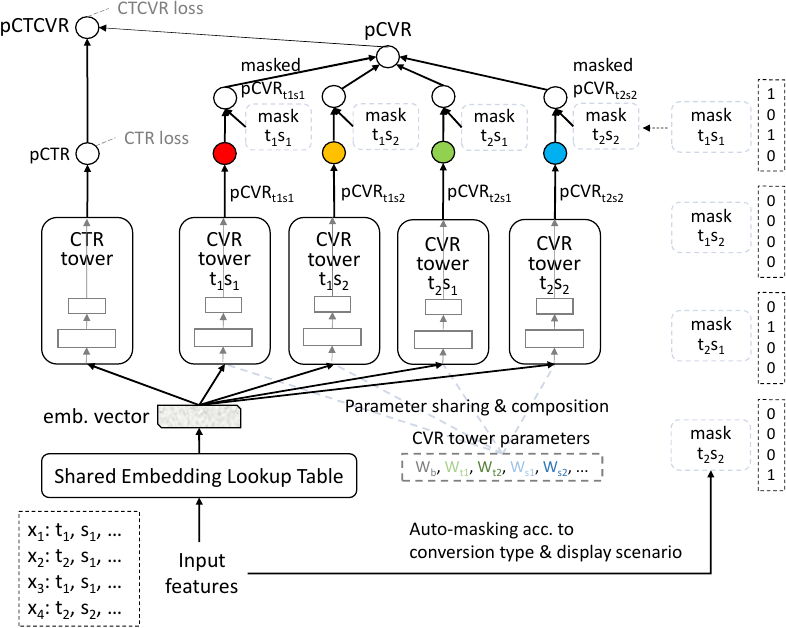}
\vskip -6pt
\caption{Masked multi-domain network. For simplicity, all the CTR and CVR prediction towers use the multi-layer perceptron consisting of several fully connected layers.}
\vskip -8pt
\label{mmn}
\end{figure*}

\subsection{Modeling Adaptive Parameters} \label{sec_param_sharing}
Figure \ref{mmn} illustrates the structure of the proposed MMN model.
It consists a CTR tower and multiple CVR towers. The inclusion of a CTR tower is to utilize the available data in the entire sample space (i.e., not only \emph{click} data but also \emph{impression} data) \cite{ma2018entire}. For simplicity, all the CTR and CVR towers use the multi-layer perceptron (MLP) consisting of several fully connected layers.
We treat each (conversion type, display scenario) pair as a domain and create a CVR tower for each domain. But we use a parameter sharing and composition strategy to reduce the number of domain-specific parameters from a product space to a sum space. At the same time, these parameters can well capture each domain's specific properties.

Given an input instance $x$, we denote the concatenated feature embedding vector after embedding lookup as $\mathbf{x}$.
We then generate the CVR prediction logit with ReLU activations \cite{nair2010rectified} as
\begin{equation}
h = MLP \Big(\mathbf{x}; \boldsymbol\theta\big(f_t(x), f_s(x)\big) \Big),
\end{equation}
where $\boldsymbol\theta$ is the set of model parameters, $f_t(x)$ is the conversion type and $f_s(x)$ is the display scenario of the input instance $x$.

\begin{figure}[!t]
\centering
\subfigure[Per-domain CVR tower parameter; per-domain dataset]{\includegraphics[width=0.44\textwidth, trim = 0 0 0 0, clip]{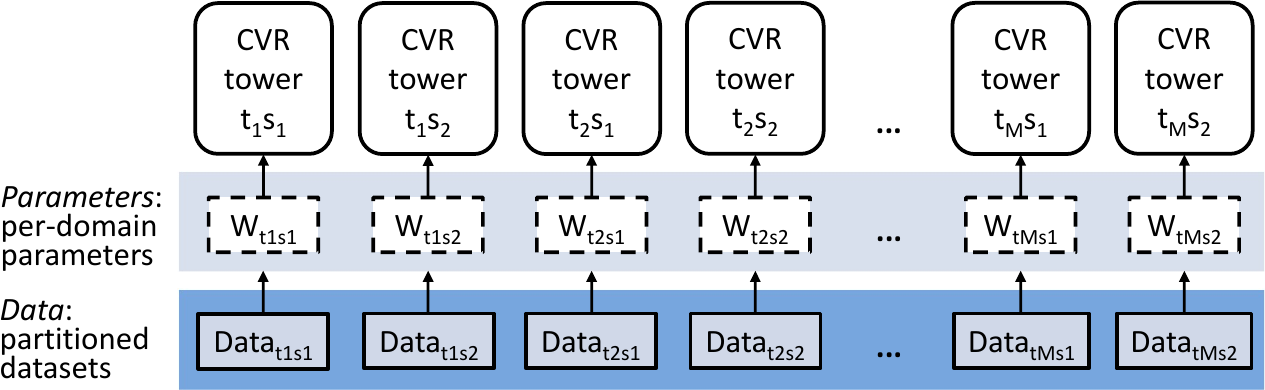}}
\vskip -2pt
\subfigure[Parameter sharing and composition; one dataset with auto-masking]{\includegraphics[width=0.44\textwidth, trim = 0 0 0 0, clip]{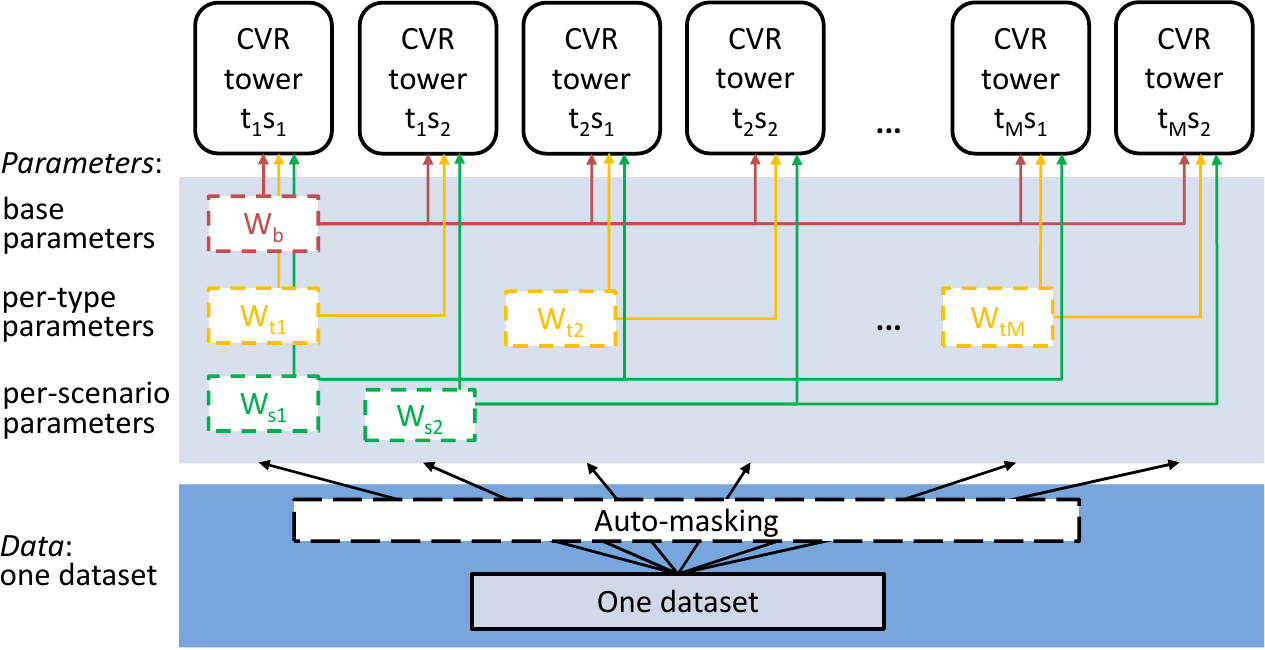}}
\vskip -8pt
\caption{(a) Per-domain CVR tower parameter; per-domain dataset. (b) Parameter sharing and composition; one dataset with auto-masking.}
\vskip -10pt
\label{param_sharing}
\end{figure}

One possible way to model $\boldsymbol\theta \big(f_t(x), f_s(x)\big)$ is to create a separate set of parameters for each (conversion type, display scenario) domain to make the model adaptive, as illustrated in Figure \ref{param_sharing}(a). Denote the number of conversion types as $N_t$ and the number of display scenarios as $N_s$. It will result in $N_t \times N_s$ sets of parameters, which makes the model too large to be trained.

In order to 1) capture commonalities across different domains, 2) reflect specific natures of different conversion types and display scenarios and 3) reduce the number of parameters, we design a parameter sharing and composition strategy as follows (illustrated in Figure \ref{param_sharing}(b)). In particular, we design
\[
\boldsymbol\theta \big(f_t(x), f_s(x)\big) =
\big\{\mathbf{W}_b^l \big\}_{l=1}^{L} + \big\{\mathbf{W}_{f_t(x)}^l \big\}_{l=1}^{L} + \big\{\mathbf{W}_{f_s(x)}^l \big\}_{l=1}^{L}.
\]
$\{\mathbf{W}_b^l\}_l$ is a set of common parameters that is shared by all the domains, in order to capture domain commonalities. $L$ is the number of fully connected layers. $\{\mathbf{W}_{f_t(x)}^l\}_l$ is a set of conversion type-specific parameters for $f_t(x)$ and $\{\mathbf{W}_{f_s(x)}^l\}_l$ is a set of display scenario-specific parameters for $f_s(x)$.
These parameters are then combined to form $\boldsymbol\theta(f_t(x), f_s(x))$.

In this way, we avoid using samples that have both type $f_t(x)$ and scenario $f_s(x)$ to train a specific set of parameters as shown in Figure \ref{param_sharing}(a), and thus mitigate the data sparsity problem. Moreover, we only create $N_t + N_s + 1$ sets of domain-specific parameters, but can generate $N_t \times N_s$ combinations to serve all different (conversion type, display scenario) domains.

After we obtain the logit $h$, the predicted CVR is given by
$
p_\textrm{cvr}(x) = \sigma(h),
$
where $\sigma$ is the sigmoid function.

\subsection{Auto-Masking} \label{sec_masking}
The above design seems fine. However, the model needs to choose specific parameters for each instance according to its conversion type and display scenario.
In modern deep learning systems such as Tensorflow \cite{abadi2016tensorflow}, instances are processed in mini-batches with the same operators.
As a result, how to process a mini-batch of instances simultaneously with the \emph{same} operators but allow each instance to use \emph{different} domain-specific parameters is still a problem.

Previous works \cite{ouyang2020minet,sheng2021one,jiang2022adaptive,cao2023towards} avoid this problem by splitting data into small per-domain datasets (Figure \ref{param_sharing}(a)). Accordingly, multiple input interfaces are created, one for each dataset.
However, applying this approach to our problem will result in $N_t \times N_s$ datasets, which incurs tremendous data processing, storage and maintenance cost.

\begin{figure*}[!t]
\centering
\includegraphics[width=0.8\textwidth, trim = 0 0 0 0, clip]{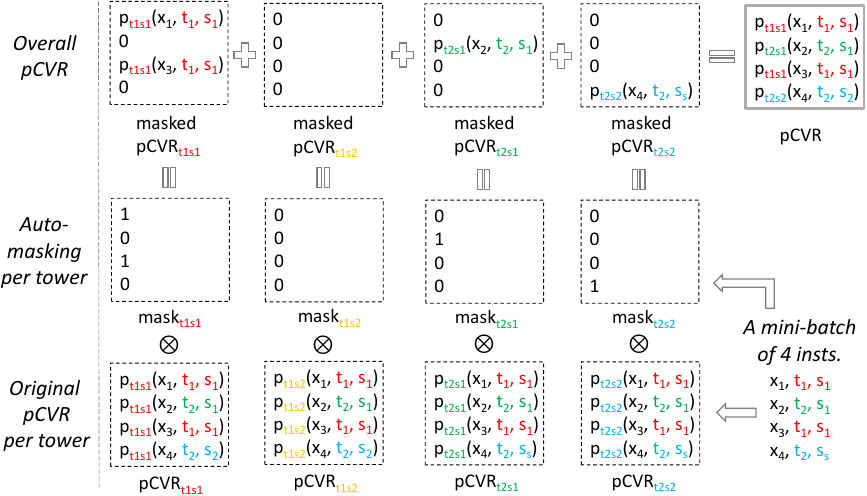}
\vskip -6pt
\caption{Illustration of how auto-masking can allow a mini-batch of instances from different (conversion type, display scenario) domains to be computed simultaneously with the same operators but different domain-specific parameters. Back propagation is fast because no gradient needs to be computed for masked entries (i.e., setting to 0).}
\vskip -8pt
\label{auto_masking}
\end{figure*}

We propose the following auto-masking strategy to solve this problem (Figure \ref{auto_masking}).
Denote a mini-batch of $N$ instances as $X = [x_1, x_2, \cdots, x_N]'$, where $'$ is the transpose operator.
We create a mask vector $\mathbf{m}_{t_i s_j}$ for each $(t_i, s_j)$ domain, where $t_i$ is the $i$th conversion type and $s_j$ is the $j$th display scenario.
The size of $\mathbf{m}_{t_i s_j}$ equals the batch size $N$.
In particular, we compute $\textbf{m}_{t_i s_j}$ as
\[
\mathbf{m}_{t_i s_j}(X) = I[f_t(X) == t_i, f_s(X) == s_j],
\]
where $I(\cdot)$ is an indication function.

We show an example in Figure \ref{auto_masking}, where we have 4 instances, $f_t(X) = [t_1, t_2, t_1, t_2]'$ and $f_s(X) = [s_1, s_1, s_1, s_2]'$.
We then have
\begin{align}
& \mathbf{m}_{t_1 s_1}(X)  = [1, 0, 1, 0]', \ \mathbf{m}_{t_1 s_2}(X)  = [0, 0, 0, 0]', \nonumber \\
& \mathbf{m}_{t_2 s_1}(X)  = [0, 1, 0, 0]', \ \mathbf{m}_{t_2 s_2}(X)  = [0, 0, 0, 1]'. \nonumber
\end{align}

The mini-batch of instances go through all the CVR towers.
Denote the output (i.e., mini-batch prediction) of the CVR tower for $(t_i, s_j)$ domain as $\mathbf{p}_{t_i s_j}(X)$.
Clearly, $X$ may contain instances that do not match $(t_i, s_j)$ and these results are incorrect.
For example, the mini-batch CVR prediction of tower $(t_1, s_1)$ in Figure \ref{auto_masking} is
\[
\mathbf{p}_{t_1 s_1}\!(X) \!=\! [p_{t_1 s_1}\!(x_1,\! t_1, \!s_1), {\color{gray} p_{t_1 s_1}\!(x_2, \! t_2, \! s_1)}, p_{t_1 s_1}\!(x_3, \! t_1, \! s_1), {\color{gray} p_{t_1 s_1}\!(x_4, \! t_2, \! s_2)}]',
\]
where gray color indicates incorrect prediction because of parameter mismatch.
We obtain the correct mini-batch CVR prediction $\mathbf{p}_\textrm{cvr}(X)$ by combining the masked outputs of all different towers
\begin{align}
& \mathbf{p}_\textrm{cvr}(X) = \sum_i \sum_j \mathbf{m}_{t_i s_j}(X) \odot \mathbf{p}_{t_i s_j}(X) \nonumber \\
= & \ [p_{t_1 s_1}(x_1, t_1, s_1), 0, p_{t_1 s_1}(x_3, t_1, s_1), 0]'
+ [0, 0, 0, 0]' + \nonumber \\
\ \ & \ [0, p_{t_2 s_1}(x_2, t_2, s_1), 0, 0]'
+ [0, 0, 0, p_{t_2 s_2}(x_4, t_2, s_2)]' \nonumber \\
= & \ [p_{t_1 s_1}(x_1, t_1, s_1), p_{t_2 s_1}(x_2, t_2, s_1), p_{t_1 s_1}(x_3, t_1, s_1), p_{t_2 s_2}(x_4, t_2, s_2)]', \nonumber
\end{align}
where $\odot$ is element-wise product.

By using the auto-masking strategy, we allow a mini-batch of instances from different (conversion type, display scenario) domains to be computed simultaneously with the same operators but adaptively with different domain-specific parameters.

\subsection{Training with Dynamically Weighted Loss} \label{sec_training}
The training of MMN is much faster than that of multi-task learning models such as MMoE \cite{ma2018modeling}, because each entry of $\mathbf{p}_\textrm{cvr}(X)$ only impacts the parameters of one tower rather than all the towers. No gradient needs to be computed during gradient back propagation for masked entries (i.e., setting to 0).

The training of the model may use the classical loss of ESMM \cite{ma2018entire}, which combines a CTR loss and a CTCVR loss as
\[
loss= loss_\textrm{ctr} + \alpha loss_\textrm{ctcvr},
\]
where $\alpha$ is a balancing parameter and
\begin{align}
& loss_\textrm{ctr} = \frac{1}{N} \sum_{n=1}^N l_\textrm{ctr}(x_n), \ loss_\textrm{ctcvr} = \frac{1}{N} \sum_{n=1}^N l_\textrm{ctcvr}(x_n), \label{loss_ori} \\
& l_\textrm{ctr}(x_n) = - \{y_n \log[p_\textrm{ctr}(x_n)] + (1-y_n) \log[1 - p_\textrm{ctr}(x_n)]\},  \nonumber\\
& l_\textrm{ctcvr}(x_n) = - \{y_n z_n \log[p_\textrm{ctr}(x_n)p_\textrm{cvr}(x_n)] \nonumber\\
& \ \ \ \ + (1 - y_n z_n) \log[1 - p_\textrm{ctr}(x_n) p_\textrm{cvr}(x_n)]\}. \nonumber
\end{align}
$N$ is the batch size, $x_n$ is the $n$th instance, $y_n$ is the click label, $z_n$ is the conversion label, and $l_\textrm{ctr}$, $l_\textrm{ctcvr}$ are cross-entropy losses.

\begin{table*}[!th]
\renewcommand{\arraystretch}{1.1}
\caption{Statistics of experimental datasets.}
\vskip -8pt
\label{tab_stat}
\centering
\begin{tabular}{|l|c|c|c|c|c|c|c|c|}
\hline
\textbf{Dataset} & \textbf{\# Fields}  & \textbf{\# Features} & \textbf{\# Train insts.} & \textbf{\# Test insts.} & \textbf{\# Conv. types (CVR range)} & \textbf{\# Display scens. (CVR range)} \\
\hline
News Feed & 58 & 61,601,087 & 153,300,184 & 5,179,132 & 19 (0.05\% $\sim$ 27.6\%) & 8 (1.6\% $\sim$ 4.1\%) \\
\hline
Criteo & 18 & 11,772,096 & 11,196,944  & 4,798,690 & 21 (6.3\% $\sim$ 18.9\%) & 17 (0.9\% $\sim$ 14.9\%) \\
\hline
\end{tabular}
\vskip -6pt
\end{table*}

Further examination on the loss reveals that it is not quite suitable for MMN. Denote the number of CVR domains as $C$, the $c$th domain as $\mathcal{D}_c$ and the number of instances that fall in the $c$th domain in a mini-batch as $N_c$. We further denote the expectation of the loss on each instance as $\bar{a}$, i.e., $E[l_\textrm{ctcvr}(x_n)] = \bar{a}$. We then have
\begin{align}
E(loss_\textrm{ctcvr}) & = \frac{1}{N} \sum_{n=1}^N E[l_\textrm{ctcvr}(x_n)]
 = \frac{1}{N} \left[ \sum_{c=1}^C \sum_{n=1}^{N_c} E[l_\textrm{ctcvr}(x_n \in \mathcal{D}_c)]  \right] \nonumber \\
& = \sum_{c=1}^C \left[\frac{1}{N}  \sum_{n=1}^{N_c} E[l_\textrm{ctcvr}(x_n \in \mathcal{D}_c)]  \right] = \sum_{c=1}^C \left(\frac{N_c}{N} \bar{a}\right). \nonumber
\end{align}
That is, the average loss applied on each domain is actually $\frac{N_c}{N} \bar{a} \leq \bar{a}$ and varies from one domain to another.
For each domain, we actually do not average over the instances corresponding to it but use an incorrect normalization value $N$.

Therefore, we propose the dynamically weighted loss as
\[
loss_\textrm{MMN} = loss_\textrm{ctr} + \alpha loss'_\textrm{ctcvr},
\]
%where
\begin{align}
loss'_\textrm{ctcvr} & = \sum_{c=1}^C \left[I(N_c > 0) \frac{1}{N_c} \sum_{n=1}^{N_c} l_\textrm{ctcvr}(x_n \in \mathcal{D}_c) \right] \nonumber \\
& = \frac{1}{N} \sum_{c=1}^C \left[I(N_c > 0) \frac{N}{N_c} \sum_{n=1}^{N_c} l_\textrm{ctcvr}(x_n \in \mathcal{D}_c) \right] \nonumber \\
& = \frac{1}{N} \sum_{n=1}^N wgt(x_n) l_\textrm{ctcvr}(x_n), \label{loss_mmn}
\end{align}
where $wgt(x_n) = \frac{N}{N_c}$ if $x_n \in \mathcal{D}_c$.

Note that the weight $wgt(x_n)$ is not pre-computed before training, but \emph{dynamically computed within each mini-batch}. This is why we call the loss as dynamically weighted loss.
In other words, even if two instances belong to the same domain, their weights could be different if they appear in different mini-batches because the number $N_c$ of samples in that domain could vary in different mini-batches.

\subsection{Fast Online Prediction}
As the online prediction is on demand, there is no need to let an online request go through all the CVR towers and use auto-masking to get the correct prediction.
Given an online request $x$ with the corresponding conversion type $f_t(x)$ and display scenario $f_s(x)$, we simply let $x$ go through the tower for $(f_t(x), f_s(x))$ to calculate a single predicted CVR, i.e., until \emph{a colored node} in Figure \ref{mmn}(a). This makes the online prediction of MMN fast.

\section{Experiments}
\subsection{Datasets} \label{sec_dataset}
The statistics of the datasets are listed in Table \ref{tab_stat}.

(1) \textbf{News feed conversion log dataset.}
This dataset contains a random sample of 14 days of ad impression, click and conversion logs from an industrial news feed advertising system for UC Toutiao.
It contains $N_t = 19$ conversion types and $N_s = 8$ display scenarios.

(2) \textbf{Criteo conversion log dataset\footnote{https://ailab.criteo.com/criteo-sponsored-search-conversion-log-dataset/}.}
This dataset contains a sample of 90 days of Criteo live traffic data \cite{tallis2018reacting}.
There are $N_t = 21$ conversion types and $N_s=17$ display scenarios.

All the features are hashed.  Each training set is further split into an initial training set (70\%) and a validation set (30\%) to find the optimal hyperparameters.

\subsection{Methods in Comparison}
We compare three categories of methods.

\emph{A. Single-task methods}
\begin{enumerate}
\item \textbf{DNN}. Deep Neural Network \cite{cheng2016wide}. It contains an embedding layer, several fully connected layers and an output layer.
\item \textbf{ESMM}. Entire Space Multi-task Model \cite{ma2018entire}. It models both CVR and CTR prediction tasks.
\end{enumerate}

\emph{B. Multi-task methods}
\begin{enumerate}
\setcounter{enumi}{2}
\item \textbf{MMoE}. Multi-gate Mixture-of-Experts model \cite{ma2018modeling}. It substitutes the shared bottom network with an MoE layer. Each task has a separate gating network.
\item \textbf{PLE}. Progressive Layered Extraction model \cite{tang2020progressive}. It separates shared components and task-specific components explicitly and adopts a progressive routing mechanism.
\end{enumerate}

\emph{C. Multi-domain methods}
\begin{enumerate}
\setcounter{enumi}{4}
\item \textbf{MT-FwFM}. Multi-Task Field-weighted Factorization Machine \cite{pan2019predicting}. As it models type-specific parameters for multi-type CVR prediction, we classify it as a multi-domain method.
\item \textbf{STAR}. Star Topology Adaptive Recommender \cite{sheng2021one}. It is proposed for CTR prediction over multiple business domains such as ``Banner'' and ``Guess What You Like''.
\item \textbf{MMN}. Masked Multi-domain Network in this paper.
\end{enumerate}

\textbf{Parameter Settings:}
We set the dimension of the embedding vectors for features as 10.
For neural network-based models, we set the number of fully connected layers in a CTR / CVR tower as $L=4$ with units \{512, 256, 128, 128\}.
All the methods are implemented in Tensorflow \cite{abadi2016tensorflow} and optimized by the Adagrad algorithm \cite{duchi2011adaptive}. We run each method 3 times and report the average results.

\subsection{Evaluation Metrics}
\textbf{Accuracy:} The Area Under the ROC Curve (AUC) is a widely used metric for CVR prediction \cite{pan2019predicting}. The larger the better.
We calculate fine-grained metrics as per conversion type AUC and per display scenario AUC. In addition, we also calculate the average AUC over all the conversion types and display scenarios.

\textbf{Scalability:} We use the number of sets of domain-specific parameters to evaluate the scalability.

\textbf{Convenience:} We use the number of separate datasets needed to train the model to evaluate the convenience.

\newcommand{\tabincell}[2]{\begin{tabular}{@{}#1@{}}#2\end{tabular}}

\begin{table*}[!t]
\setlength{\tabcolsep}{2pt}
\renewcommand{\arraystretch}{1}
\caption{Test AUCs on the \emph{News Feed} ad dataset. The best result is in bold font. Randomly selected conversion types / display scenarios that span the training CVR spectrum are shown. (type / scen.): a method optimizes only the type / scenario dimension because of the out-of-memory issue. A small improvement in AUC (e.g., 0.0020) can lead to a significant increase in online CVR.}
\vskip -8pt
\label{tab_auc_news_feed}
\centering
\begin{tabular}{|c|l|c|c|c|c|c|c|c|c||c|c|c|c||c|}
\hline
\multicolumn{2}{|c}{} & \multicolumn{8}{|c||}{\textbf{Conversion types}} & \multicolumn{4}{|c||}{\textbf{Display scenarios}} & \\
\hline
& Type / Scen. code & 50 & 44 & 65 & 18 & 1 & 11 & 5 & 66         & 3126 & 8082 & 5739 & 4661    & Avg. \\
\hline
Method & CVR & 27.6\% & 14.9\% & 3.1\% & 2.8\% & 1.3\% & 0.5\% &0.3\% & 0.05\%        &4.1\% & 3.7\% & 2.1\% & 1.6\% & \\
\hline
\multirow{2}{*}{\tabincell{c}{Single-\\task}}
&DNN &0.8584 &0.7247 &0.6424 &0.7898 &0.8402 &0.7941 &0.7815 &0.6825      &0.9688 &0.8112 &0.9346 &0.9698    &0.8165 \\
&ESMM &0.8576 &0.7253 &0.6519 &0.7906 &0.8404 &0.7927 &0.7852 &0.6829     &0.9687 &0.8114 &0.9343 &0.9697    &0.8176 \\
\hline
\multirow{4}{*}{\tabincell{c}{Multi-\\task}}
&MMoE (type) &0.8574 &0.7157 &0.6565 &0.7953 &0.8406 &0.7909 &0.7861 &0.6796    &0.9686 &0.8110 &0.9340 &0.9698    &0.8171 \\
&MMoE (scen.) &0.8574 &0.6994 &0.6449 &0.7921 &0.8385 &0.7908 &0.7824 &0.6792    &0.9683 &0.8110 &0.9330 &0.9693   &0.8139 \\
&PLE (type) &0.8578 &0.7258 &0.6499 &0.7948 &0.8411 &0.7916 &0.7865 &0.6832    &0.9685 &0.8116 &0.9346 &0.9698    &0.8179 \\
&PLE (scen.) &0.8577 &0.7231 &0.6525 &0.7931 &0.8402 &0.7931 &0.7852 &0.6976    &0.9687 &0.8118 &0.9342 &0.9701    &0.8189 \\
\hline
\multirow{4}{*}{\tabincell{c}{Multi-\\domain}}
&MT-FwFM &0.8508 &0.6670 &0.6273 &0.7764 &0.8272 &0.7648 &0.7585 &0.5835    &0.9676 &0.8004 &0.9288 &0.9663    &0.7932  \\
&STAR (type) &0.8616 &0.7554 &0.6687 &0.7832 &0.8419 &0.7952 &0.7865 &0.7018    &0.9691 &0.8125 &0.9353 &0.9703    &0.8235 \\
&STAR (scen.) &0.8583 &0.7550 &0.6642 &0.7924 &0.8417 &0.7946 &\textbf{0.7868} &0.6677    &0.9691 &0.8148 &0.9380 &0.9703    &0.8211 \\
&MMN &\textbf{0.8626} &\textbf{0.7564} &\textbf{0.6842} &\textbf{0.7962} &\textbf{0.8427} &\textbf{0.8063} &0.7865 &\textbf{0.7219}     &\textbf{0.9710} &\textbf{0.8168} &\textbf{0.9411} &\textbf{0.9732}    &\textbf{0.8299} \\
\hline
\end{tabular}
%\vskip -8pt
\end{table*}

\begin{table*}[!t]
\setlength{\tabcolsep}{2pt}
\renewcommand{\arraystretch}{1}
\caption{Test AUCs on the \emph{Criteo} ad dataset. The best result is in bold font.}
\vskip -8pt
\label{tab_auc_criteo}
\centering
\begin{tabular}{|c|l|c|c|c|c|c|c||c|c|c|c|c|c||c|}
\hline
\multicolumn{2}{|c}{} & \multicolumn{6}{|c||}{\textbf{Conversion types}} & \multicolumn{6}{|c||}{\textbf{Display scenarios}} & \\
\hline
& Type / Scen. code & 7647 & 1208 & 4445 & 6198 & 7652 & 1584         & 5264 & 8441 & 1132 & 9036 & 1438 & 8972  & Avg. \\
\hline
Method & CVR & 18.9\% & 13.2\% & 11.4\% & 9.6\% & 8.0\% & 6.3\%        &14.9\% & 13.5\% & 12.0\% & 10.2\% & 6.4\% & 0.9\% & \\
\hline
\multirow{2}{*}{\tabincell{c}{Single-\\task}}
&DNN &0.7606 &0.7556 &0.7638 &0.7762 &0.8220 &0.7376        &0.7310 &0.6383 &\textbf{0.6191} &0.7402 &0.6579 &0.7708     &0.7311 \\
&ESMM & 0.7593 & 0.7538 & 0.7645 & 0.7707 & 0.8213 & 0.7405       &0.7310 &0.6310 & 0.6156 &0.7365 &0.6547 &0.7792   &0.7298 \\
\hline
\multirow{4}{*}{\tabincell{c}{Multi-\\task}}
&MMoE (type) &0.7129 & 0.7484 & 0.7556 & 0.7681 & 0.8176 & 0.5444    &0.7028 &0.6212 &0.6089 &0.7030 &0.5727 &0.7359   &0.6910 \\
&MMoE (scen.) &0.7659 & 0.7463 &0.7460 & 0.7686 &0.8168 & 0.7238      &0.7239 &0.5469 &0.5032 &0.7162 &0.5033 &0.7596    &0.6934 \\
&PLE (type) &0.7468 &0.7512 &0.7593 &0.7729 &0.8212 &0.7323      &0.7267 &0.6297 &0.6123 &0.7371 &0.6511 &0.7683     &0.7257 \\
&PLE (scen.) &0.7565 &0.7526 &0.7606 &0.7731 &0.8224 & 0.7372     &0.7255 &0.6321 &0.6105 &0.7381 &0.6432  &0.7752    &0.7273 \\
\hline
\multirow{4}{*}{\tabincell{c}{Multi-\\domain}}
&MT-FwFM &0.7491 &0.7345 &0.7294 &0.7528 &0.8034 &0.7112      & 0.6831 &0.6078 &0.5858 &0.6788 &0.5953 &0.7239   &0.6963 \\
&STAR (type) & 0.7692 & 0.7585 & 0.7663 & 0.7776 & 0.8237 & 0.7445        &0.7340 &0.6547 &0.6176 &0.7356 &0.6648 &0.7844    &0.7359 \\
&STAR (scen.) & 0.7715 & 0.7565 & 0.7644 & 0.7680 & 0.8226 & 0.7458       &0.7352 &\textbf{0.6611} &0.6155 &0.7373 &0.6614 &0.7878    &0.7356 \\
&MMN & \textbf{0.7741} & \textbf{0.7586} & \textbf{0.7689} & \textbf{0.7834} & \textbf{0.8251} & \textbf{0.7473}                   &\textbf{0.7370} &0.6569 &0.6158 &\textbf{0.7413} &\textbf{0.6747} &\textbf{0.7895}    &\textbf{0.7394} \\
\hline
\end{tabular}
%\vskip -8pt
\end{table*}

\subsection{Offline Performance}
\subsubsection{\textbf{Accuracy.}}
Table \ref{tab_auc_news_feed} and Table \ref{tab_auc_criteo} list the test AUCs. As multi-task methods are not quite suitable for our problem setting, MMoE and PLE perform even worse than single-task methods such as DNN and ESMM on the Criteo dataset. On the News Feed dataset, their performance is comparable.
Multi-domain method STAR outperforms single-task methods such as DNN and ESMM.
MMN achieves the highest AUC in most cases. It is because MMN has the ability to tackle both conversion type and display scenario simultaneously by design. In contrast, other multi-task / multi-domain methods can either optimize on the flattened type $\times$ scenario space with significant data sparsity issue, or optimize only one dimension because of the out-of-memory issue.

\subsubsection{\textbf{Scalability.}}
We use the number of sets of domain-specific parameters to evaluate the scalability.
Single-task methods do not model domain-specific parameters and do not have the scalability issue.
MMN models $N_t + N_s + 1$ sets while other multi-task / multi-domain methods model $N_t \times N_s$ sets of domain-specific parameters. Compared with these methods, MMN reduces the number of sets of domain-specific parameters from 357 to 39 (i.e., 89.1\% reduction) in the Criteo dataset and from 152 to 28 (i.e., 81.6\% reduction) in the News Feed dataset.

\subsubsection{\textbf{Convenience.}}
We use the number of separate datasets needed to train the model to evaluate the convenience.
Single-task / multi-task methods use only one dataset and they are convenient. With the help of auto-masking, MMN also uses only one dataset. However, other multi-domain methods need to create a separate dataset for each domain and results in $N_t \times N_s$ sub datasets.
Compared with these methods, MMN reduces the number of separate datasets from 357 to 1 (i.e., 99.7\% reduction) in the Criteo dataset and from 152 to 1 (i.e., 99.3\% reduction) in the News Feed dataset.

\subsubsection{\textbf{Ablation Study: effect of type-specific and scenario-specific parameters.}}
\begin{figure}[!t]
\centering
\subfigure[News Feed: AUC(MMN) - AUC\big(MMN (common params.)\big)]{\includegraphics[width=0.45\textwidth, trim = 16 20 12 10, clip]{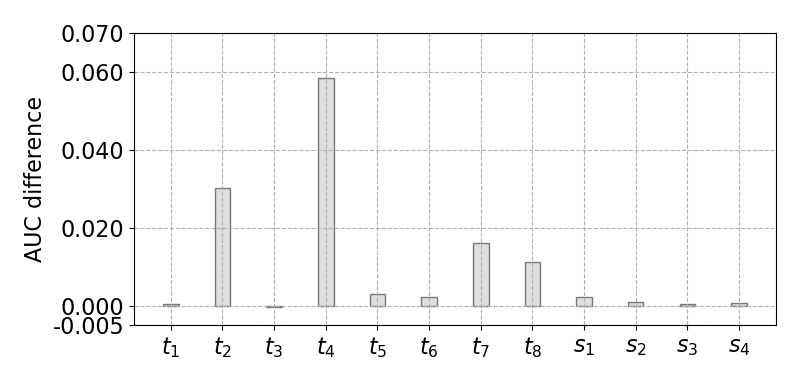}}
%\vskip -4pt
\subfigure[Criteo: AUC(MMN) - AUC\big(MMN (common params.)\big)]{\includegraphics[width=0.45\textwidth, trim = 16 20 12 10, clip]{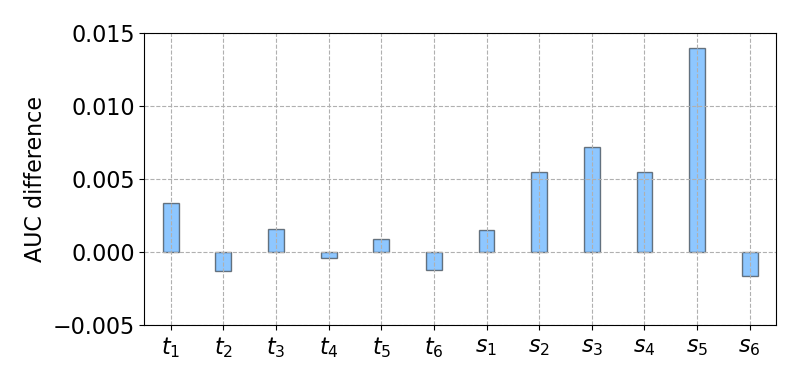}}
\vskip -6pt
\caption{Effect of domain-specific parameters.}
\vskip -6pt
\label{fig_param}
\end{figure}

Figure \ref{fig_param} plots the AUC difference between MMN and MMN (common params.).
In MMN (common params.), we only use shared common parameters $\{\mathbf{W}_b^l\}_l$ for all the domains. In other words, no type-specific and scenario-specific parameters are learned.
It is observed that MMN outperforms MMN (common params.) in most cases. These results show that different conversion types and display scenarios do have their own properties and using the same parameter set cannot well capture individual properties.

\subsubsection{\textbf{Ablation Study: effect of dynamically weighted loss.}}
\begin{figure}[!t]
\centering
\subfigure[News Feed: AUC(MMN) - AUC\big(MMN (no dynamic weight)\big)]{\includegraphics[width=0.45\textwidth, trim = 16 20 12 12, clip]{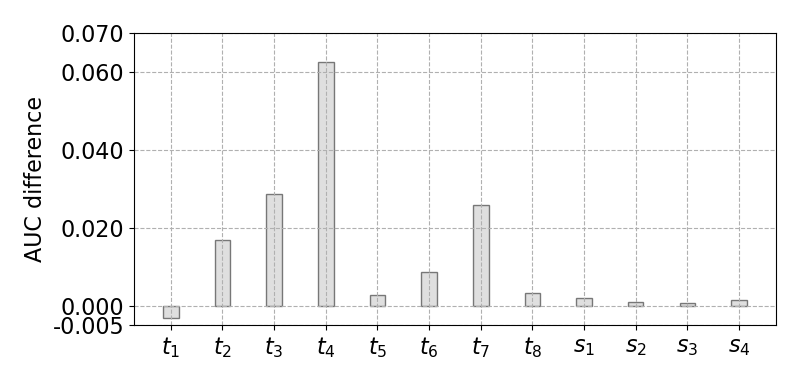}}
%\vskip -4pt
\subfigure[Criteo: AUC(MMN) - AUC\big(MMN (no dynamic weight)\big)]{\includegraphics[width=0.45\textwidth, trim = 16 20 12 12, clip]{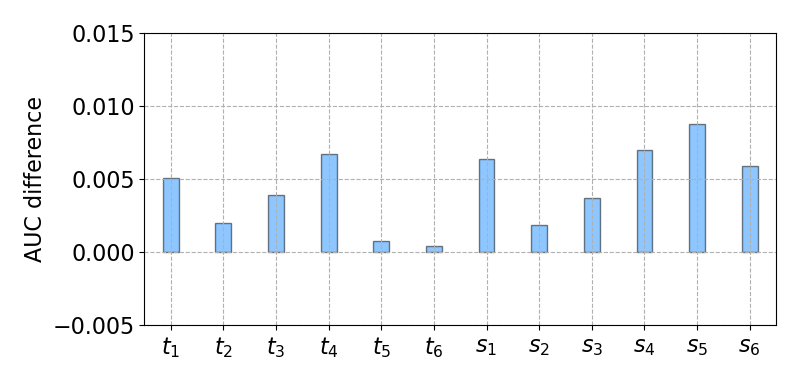}}
\vskip -6pt
\caption{Effect of dynamically weighted loss.}
\vskip -6pt
\label{fig_loss}
\end{figure}

Figure \ref{fig_loss} plots the AUC difference between MMN and MMN (no dynamic weight).
In the latter, we use the loss in Eq.(\ref{loss_ori}) for model training.
It is observed that AUCs of MMN improve a lot compared with MMN (no dynamic weight) in most cases. These results validate the effectiveness of the dynamically weighted loss.

\subsection{Online Deployment}
We deployed MMN in an industrial news feed advertising system for UC Toutiao which serves hundreds of millions of traffic every day.
MMN is trained using a distributed CPU cluster which contains 8 servers each with 96 CPU core units and 512 GB RAM.
By design, MMN can satisfy the scalability and convenience requirements. In practice, we do not encounter the out-of-memory issue when training MMN but when training STAR on the type $\times$ scenario space. Moreover, MMN needs only one dataset to train the model, and it significantly reduces the data processing overhead.
We conducted online experiments in an A/B test framework over 14 days in August 2022, where the base serving model is ESMM.
By design, the online prediction time of MMN is short. In practice, the prediction time of MMN is 2.3 ms, which is only slightly longer than that of ESMM, which is 2.2 ms.
MMN leads to clear online CVR increases in all the 19 conversion types and the 8 display scenarios, where the increases range from 3.2\% to 12.6\%. Moreover, the CPM (revenue) is increased by 5.1\%.
After the A/B test, MMN provides real-time CVR prediction for all ad traffic in our system.

\section{Related Work}
\textbf{CVR prediction.}
The task of CVR prediction \cite{lee2012estimating,rosales2012post,chapelle2014modeling,lu2017practical} in online advertising is to estimate the probability of a user makes a conversion event on a specific ad after a click event.
\cite{lee2012estimating} estimates CVR based on past performance observations along data hierarchies. \cite{chapelle2014simple} proposes an LR model for CVR prediction and \cite{agarwal2010estimating} proposes a log-linear model.
\cite{ma2018entire} proposes the ESMM model to jointly consider CVR and CTR prediction tasks and exploit data in the entire sample space.
ESM$^2$ \cite{wen2020entire}, GMCM \cite{bao2020gmcm} and HM$^3$ \cite{wen2021hierarchically} exploit additional purchase-related behaviors after click (e.g., favorite, add to cart and read reviews) for CVR prediction. But such behaviors are not always available in different advertising systems.

\textbf{Multi-task/multi-domain learning.}
In multi-task learning \cite{ruder2017overview,misra2016cross,liu2023multi}, the domains are the same and the tasks are different. In multi-domain learning \cite{dredze2010multi,joshi2012multi,yang2022adasparse}, the domains are different and the tasks are the same.
Because of such differences, existing multi-task learning models such as Shared-Bottom, OMoE, MMoE, CGC and PLE \cite{ma2018modeling,zhao2019recommending,ma2019snr,tang2020progressive} cannot well address our problem.
MT-FwFM \cite{pan2019predicting}, STAR \cite{sheng2021one} and ADIN \cite{jiang2022adaptive} are multi-domain models that are more closely related to our work. However, they cannot address both different conversion types and different display scenarios. Moreover, they are not designed to satisfy the scalability and convenience requirements for our problem.

\section{Conclusion}
In this paper, we present the design and evaluation of masked multi-domain network (MMN) to address the multi-type and multi-scenario CVR prediction problem in a real-world advertising system. We design various strategies such as modeling domain-specific parameters, parameter sharing and composition, auto-masking and dynamically weighted loss in order that MMN can simultaneously satisfy the accuracy, scalability and convenience requirements as an industrial model. Experimental results show that MMN does achieve these requirements and it outperforms several state-of-the-art models for multi-type and multi-scenario CVR prediction.

\bibliographystyle{ACM-Reference-Format}
\balance
\bibliography{ref}

\end{document}